\documentclass[aps,twocolumn,floats,superscriptaddress,showpacs]{revtex4}

\usepackage{graphicx}

\begin{document}

\title{
Extended Dynamical Mean Field Theory Study of the Periodic Anderson Model
}
\author{Ping Sun}
\author{Gabriel Kotliar}
\affiliation{
Center for Materials Theory,
Department of Physics and Astronomy,
Rutgers University,
Piscataway, NJ 08854-8019}
\date{\today}

\begin{abstract}
We investigate the competition of the Kondo and the RKKY interactions
in heavy fermion systems. We solve a periodic Anderson model using
Extended Dynamical Mean Field Theory (EDMFT) with QMC. We monitor
simultaneously the evolution of the electronic and magnetic
properties. As the RKKY coupling increases the heavy fermion
quasiparticle unbinds and a local moment forms. At a critical RKKY
coupling there is an onset of magnetic order. Within EDMFT the two
transitions occur at different points and the disapparence of the
magnetism is not described by a local quantum critical point.
\end{abstract}


\maketitle

The interaction between local moments and conduction electrons is a long
standing problem in condensed matter physics. As discussed early on
\cite{doniach,varma}, two competing interactions, the Kondo exchange
between the conduction electrons and the moments, and the indirect exchange
among the moments mediated by the RKKY interaction, are central to this
problem.

The physics at high temperatures is well understood. Moments and
conduction electrons retain their identities and interact weakly with
each other. The central question in this field is to understand the
characteristics of the state that develops as the temperature is
lowered. When the RKKY interaction is much larger than the Kondo
energy, the moments order magnetically and the conduction electrons
follow their magnetization without undergoing a strong
renormalizations.  The spin entropy is mostly quenched by the ordering
of the local moments. When the RKKY interaction is much weaker than
the Kondo energy, the spin entropy is absorbed by the interaction of
the moments and the conduction electrons. This results in the
formation of a heavy Fermi liquid of quasiparticles which are
composites of local moment spins bound to the conduction electrons.
The description of the low temperature state, when both the RKKY and
the Kondo interactions are of comparable magnitude, has been an
important unresolved question. It has received renewed theoretical
interest motivated by the intensive experimental investigation
\cite{stewart} of two materials, YbRh$_2$Si$_2$ \cite{YbRh} and
CeCu$_{6-x}$Au$_x$ \cite{CeCu}, which can be driven continuously from
the paramagnetic (PM) phase to the antiferromagnetic (AF) phase by
application of pressure, alloying and magnetic field. Near the quantum
critical point, RKKY and Kondo interactions are both essential and
need to be treated on the same footing. It has not been possible
\cite{coleman0,coleman1} to describe the quantum phase transition in
these materials within the standard Hertz-Millis-Moriya (HMM)
framework \cite{hertz,millis,moriya} so far.

The interplay of Kondo and RKKY interactions has been addressed
using slave boson mean field methods and large-N expansions \cite{largen}.
This method can capture the Kondo effect and the Fermi liquid
phase. The introduction of bond variables allows some
description of the effects of magnetic correlations in the
heavy Fermi liquid state. However the simultaneous description of the
antiferromagnetism and heavy Fermi liquid behavior within this technique
is still an open problem. Furthermore, mean field methods can not
capture the incoherent part of the electron Green's function which
is important near the transition. This problem becomes more tractable
near the PM metal to spin glass transition \cite{sengupta}.

With the development of Dynamical Mean Field Theory (DMFT)
\cite{gabi1}, which can treat both the AF states and the Kondo effect,
it has been possible to make further progress in this problem
\cite{jarrell,rozenberg}. An extension of DMFT (EDMFT) \cite{subir},
which allows a better treatment of the competition between the
exchange interaction and kinetic energy, was first applied in the
context of a one-band model by Parcollet and Georges \cite{olivier}
and more recently by Haule {\it et al.} \cite{haule} who established
that the increasing exchange reduces the coherence temperature.  The
Kondo lattice with magnetic frustration in the large-N limit was
recently studied by Burdin {\it et al.} \cite{burdin} who found a
paramagnet to spin liquid quantum phase transition.  Si {\it et al.}
\cite{si0} considered the $SU(2)$ Kondo lattice model in the view
point that the model can be described in terms of the criticality of
an impurity model in a self-consistent medium. They concluded that in
2D and within EDMFT this model a) has a quantum critical point, b) the
quantum critical point has nonuniversal exponents, and c) a numerical
calculation \cite{si1} of this exponent has remarkable similarity to
some of the experimental observations in the CeCu$_{6-x}$Au$_x$
system. However, no study of the magnetism had been carried out and
this is the subject of this paper.

Our goal is to address these issues within the periodic Anderson model
(PAM), focusing more on the evolution of the electronic structure at
finite temperatures and not too close to the phase transition. We
solve the model using 
EDMFT and a continuous field Hubbard-Stratonovich QMC method as an
impurity solver \cite{motome,ping}.

The PAM Hamiltonian is given by:

\[
  H=\sum_{\vec{k}\sigma} \epsilon_{\vec{k}} n^c_{\vec{k}\sigma}
  +V\sum_{i\sigma}(c_{i\sigma}^{\dagger}f_{i\sigma}
  +f_{i\sigma}^{\dagger}c_{i\sigma})
  +E_f\sum_{i\sigma}n^f_{i\sigma}
\]
\begin{equation}
\label{eq-01}
  +U\sum_{i}\left(n^f_{i\uparrow}-1/2\right)
  \left(n^f_{i\downarrow}-1/2\right)+\frac{J}{2}\sum_{\langle ij \rangle}
  S^f_{i,z} S^f_{j,z}
\end{equation}

\noindent where $c_{i\sigma}$ ($f_{i\sigma}$) annihilates a conduction
(localized f-) electron of spin $\sigma$ at site $i$.
$n^a_{i\sigma}=a_{i\sigma}^{\dagger}a_{i\sigma}$, with $a=c,f$, and
$S^f_{i,z}=n^f_{i\uparrow}-n^f_{i\downarrow}$.  We introduce an
independent RKKY interaction which can be induced by hybridization of
the f-electron with either the c-electrons or the electrons in the
other orbitals which are not included explicitly \cite{gabi0}. There
is experimental evidence that the spin fluctuations
\cite{rosch,stockert} near the critical doping are of quasi-2D nature
in CeCu$_{6-x}$Au$_x$. For technical convenience, we take a short
range Ising AF RKKY exchange of the form $J_{\vec{q}}=J_{RKKY}(\cos
q_x +\cos q_y)/2$. For the c-band we take the dispersion,
$\epsilon_{\vec{k}}=(\cos k_x +\cos k_y+\cos k_z)/3$.

We then make the EDMFT approximation \cite{subir,motome,ping}.  To
describe the AF phase \cite{gabi1} we introduce formally two effective
impurity models and use the symmetry that electrons at one impurity
site is equivalent to the electrons on the other with opposite spins.
The self-consistent condition involves the self-energies of both
spins. The EDMFT equations in the PM phase are obtained when the
self-energies do not depend on spin.  The local Green's functions and
the self-energies are obtained from the quantum impurity model:

\[
  S_0=-\int_0^{\beta} d\tau \int_0^{\beta} d\tau^{\prime} \sum_{\sigma}
  f_{0,\sigma}^{\dagger}(\tau) {\cal G}_{0,\sigma}^{-1}
  (\tau-\tau^{\prime}) f_{0,\sigma}(\tau^{\prime})
\]
\[
  -\frac{1}{2}\int_0^{\beta} d\tau \int_0^{\beta} d\tau^{\prime}
  \phi_{0}(\tau) {\cal D}_{0}^{-1}(\tau-\tau^{\prime})
  \phi_{0}(\tau^{\prime})
\]
\[
  +U \int_0^{\beta} d\tau 
  \left[n^f_{0,\uparrow}(\tau)-1/2\right]
  \left[n^f_{0,\downarrow}(\tau)-1/2\right]
\]
\begin{equation}
\label{eq-02}
  -\int_0^{\beta} d\tau \phi_{0}(\tau)\left[n^f_{0,\uparrow}(\tau)
  -n^f_{0,\downarrow}(\tau)\right].
\end{equation}

\noindent ${\cal G}_0$ and ${\cal D}_0$ play a role of the dynamical
Weiss functions and describe the effect of the environment on the f
electron and its spin. Their behavior is obtained from a
self-consistent solution of the EDMFT equations:

\begin{equation}
\label{eq-03}
  G^{imp}_{ff,\sigma}(ip_n)
  =\sum_{\vec{k}} \frac{\xi_{\sigma}(k,ip_n)}{
   [\xi_{\sigma}(\vec{k},ip_n)][\xi_{-\sigma}(\vec{k},ip_n)]
   -\eta(\vec{k},ip_n)^2},
\end{equation}

\begin{equation}
\label{eq-04}
  D^{imp}(i\omega_n)=\sum_{\vec{q}} J_{\vec{q}}/[1-J_{\vec{q}}
  \Pi(i\omega_n)].
\end{equation}

\noindent In Eq.\ref{eq-03} we used

\[
  \xi_{\sigma}(\vec{k},ip_n)=ip_n+\mu
  -\sigma[J_{\vec{q}=0}+{\cal D}_0(i0)]S_z
\]
\[
  -E_f-\Sigma_{-\sigma}(ip_n)
  -V^2\frac{ip_n+\mu}{(ip_n+\mu)^2-\epsilon_{\vec{k}}^2},
\]

\noindent $\eta(\vec{k},ip_n)=V^2\epsilon_{\vec{k}}/
[(ip_n+\mu)^2-\epsilon_{\vec{k}}^2]$, and
$S_z=G^{imp}_{ff,\uparrow}(0^-)-G^{imp}_{ff,\downarrow}(0^-)$.  The
f-electron and the boson self-energies are obtained by the local Dyson
equations. $p_n$ and $\omega_n$ are the fermion and boson Matsubara
frequencies, respectively. The boson self-energy and the
experimentally relevant spin susceptibility are related by
\cite{ping}:

\begin{equation}
\label{eq-05}
  [\chi(\vec{q},i\omega_n)]^{-1} =
  -J_{\vec{q}}+[\Pi(i\omega_n)]^{-1}
\end{equation}

We use QMC \cite{ping} to solve the impurity problem described by
Eq.(\ref{eq-02}). Eqs.(\ref{eq-03}) and (\ref{eq-04}) are then
applied.  This forms one iteration loop from which the new dynamical
Weiss functions are obtained for the next iteration until convergence.
We use $U=3.0$, $V=0.6$, $E_f=-0.5$, and $\mu=0.0$. The number of time
slices in QMC is $L=16$ for $\beta\le 8.0$, $L=32$ for $8.0\le \beta
\le 20.0$, and $L=64$ for $16.0\le \beta \le 32.0$. The number of QMC
sweeps is typically $10^6$. We measure the energies in terms of the
Kondo temperature at zero $J_{RKKY}$ coupling,
$T_k^0=1/\beta_k^0\simeq 1/8.0$, which is determined from the position
of the peak in the local susceptibility vs temperature plot (see
Fig. \ref{fig-03} below).

\begin{figure}[ht]
\includegraphics[width=8cm]{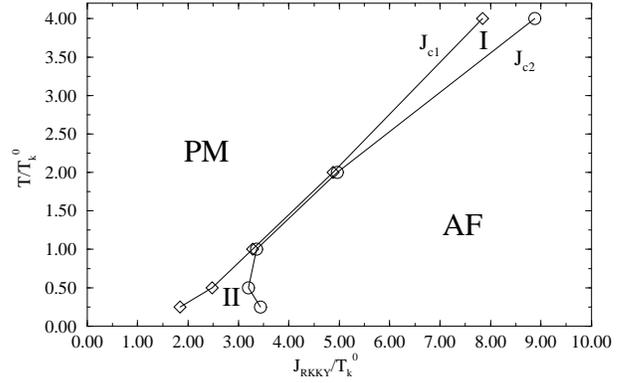}
\caption{The EDMFT phase diagram of the PAM.
The diamonds and circles are the EDMFT results of the phase boundaries.
In between the two solid lines connecting the symbols,
$J_{c1}$ and $J_{c2}$, we find both the PM and the AF solutions.}
\label{fig-01}
\end{figure}

Fig. \ref{fig-01} describes the EDMFT phase diagram of the PAM as
temperature and $J_{RKKY}$ are varied. Within EDMFT the PAM exhibits a
first order phase transition. There are two lines, $J_{c1}$ and
$J_{c2}$, betweens which we find two EDMFT solutions.  The strength of
the first order phase transition, which is reflected in the size of
the coexistence region, is non-monotonic in temperature. It is large
at high (region I) and low (region II) temperatures. At the boundaries
of the coexistence region, the magnetization $S_z$ and the static
local susceptibility are discontinuous. Along the line $J_{c2}$ the
transition becomes stronger first order as the temperature is lowered,
while along the $J_{c1}$ line it becomes weaker \cite{ping1}. The
overall transition tends to be more strongly first order at low
temperatures. As we increase $J_{RKKY}$ at fixed $T$, the f-electron
becomes more localized (${\rm Im}[{\cal G}_0^{-1} (ip_n)]-p_n$ becomes
smaller at low frequencies) while the local spin fluctuations are
strongly enhanced in the neighborhood of the $J_{c2}$ line (${\cal
D}_0(i\omega_n)$ becomes bigger at low frequencies)
\cite{ping1}. EDMFT self-consistency is crucial in obtaining the
results, which are absent in the simple impurity model.

\begin{figure}[ht]
\includegraphics[width=8cm]{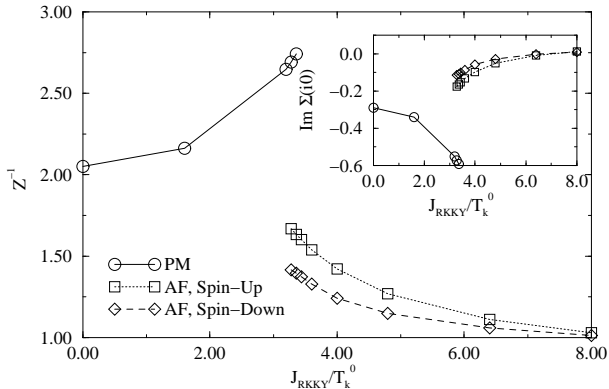}
\caption{The evolution of the inverse quasiparticle residue $Z^{-1}$
at a fixed temperature $T/T_k^0=1.0$ as a function $J_{RKKY}$. In our
calculation, the majority spin is spin-up. The majority spin band in
the AF phase is more strongly renormalized. In the inset, we plot the
extrapolation of the f-electron self-energy to zero Matsubara
frequency. The bigger this value, the shorter the lifetime of the
excitations.}
\label{fig-02}
\end{figure}

An important question is the effect of the RKKY interaction on the
quasiparticle mass. The RKKY interaction partially locks the spins and
thus reduces the effective mass and the spin entropy. But it also
serves as an additional interaction among the quasiparticles which
increases the effective mass. To address this question, we plot at
$T/T_k^0=1.0$ the inverse quasiparticle residue $Z^{-1}$ as a function
of $J_{RKKY}$ in Fig. \ref{fig-02}.  $Z^{-1}$, which is proportional
to the effective electron mass, gets enhanced as the transition is
approached from both sides.  This mass enhancement is not included in
the HMM. We predict that the mass enhancement of the majority carriers
is larger than that of the minority carriers in the AF phase. In the
inset to Fig. \ref{fig-02} we display the extrapolation of the
self-energy to zero Matsubara frequency which is a measure of the
quasiparticle lifetime. Notice that in the PM side, the inelastic
effects increase as the transition is approached, an effect which is
also absent in weak coupling approaches. This trend was observed in
earlier EDMFT studies \cite{olivier,haule}.  Given the parameters we
use, the f-electron is delocalized \cite{ping1} throughout the phase
diagram. However, as $J_{RKKY}$ is increased, the f-electron becomes
more localized.

\begin{figure}[ht]
\includegraphics[width=8cm]{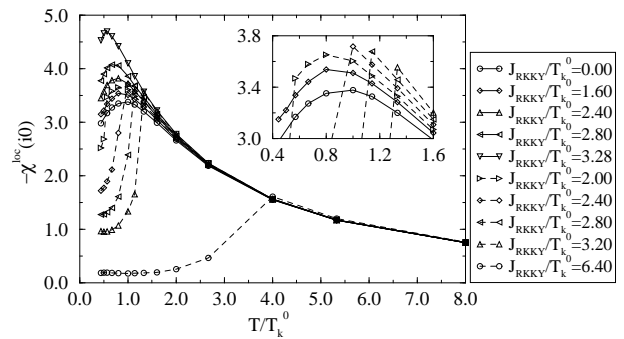}
\caption{ The evolution of Kondo peak into the Neel cusp as $J_{RKKY}$
is increased. In the cases where the symbols are connected by the
solid lines, we solve the EDMFT equations by forcing the PM order. The
results with dashed lines are obtained without such a constraint. The
inset shows the behavior around the Kondo peak with the same symbol
scheme.}
\label{fig-03}
\end{figure}

We now turn to the evolution of the magnetic properties as a function
of temperature and $J_{RKKY}$ . In Fig. \ref{fig-03} we plot the local
susceptibility. Following the PM solution we sweep the phase diagram
from $J_{RKKY}=0$ up to the line $J_{c2}$ in Fig. \ref{fig-01}. In
this procedure, the position of the Kondo peak moves towards lower
temperature and its height has an increment of about $37\%$. If we
solve the EDMFT equations without forcing any magnetic order, either
the PM or AF solution can appear.  For $J_{RKKY}/T_k^0 \ge 2.00$, we
see the Neel cusp first occurs below the Kondo peak which disappears
as $J_{RKKY}$ increases further. This happens before the Kondo
temperature goes to zero, consistent with the HMM.

\begin{figure}[ht]
\includegraphics[width=8cm]{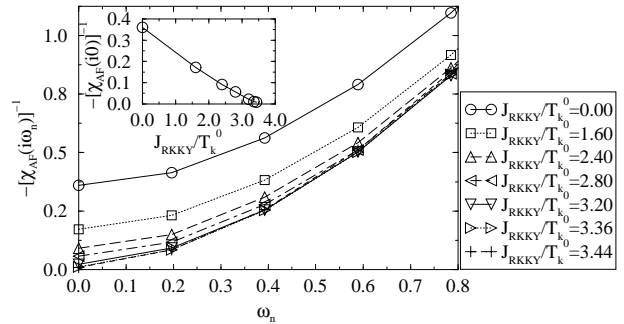}
\caption{The inverse static spin susceptibility at the AF ordering 
wave vector
$\vec{k}=(\pi,\pi)$ at $T/T_k^0=0.25$. We plot the frequencies $\omega_n\le
2\pi/\beta_k^0$. The inset is the susceptibility at
zero Matsubara frequency vs the RKKY exchange.
}
\label{fig-04}
\end{figure}

To further test the local critical scenario \cite{si0} we follow the
evolution of the PM solution as a function of temperature and
frequency.  In Fig. \ref{fig-04} we plot $-1/\chi_{AF}(i\omega_n)$,
which is defined as $-1/\chi(\vec{q}, i\omega_n)$ at
$\vec{q}=(\pi,\pi)$, as $J_{RKKY}$ is changed. One can see from the
inset that, as the transition is approached, the zero frequency value
$1/\chi_{AF}(i0)$ vanishes. From the curvature of the lines one can
see that the frequency dependences of $1/\chi_{AF}(i\omega_n)$ does
not seem to be consistent with a sublinear behavior required by the
local quantum critical scenario. In stead, it is compatible with the
standard HMM picture and a recent large-N study \cite{pankov1}.

In summary we have studied the PAM within EDMFT. In the parameter
regime studied, we found that the heavy quasiparticles first form, and
then undergo a magnetic phase transition.  This would indicate that
the HMM description of the quantum phase transition would apply to
this model. On the other hand, we also find a large enhancement of
the effective mass and reduction of the quasiparticle lifetime, which
are beyond the standard model.

This paper raises a large number of questions that require further
investigations. The first one is the connection between our results
and those reported in EDMFT studies of the Kondo lattice models by Si
{\it et al.} \cite{si0,si1,si2}. Our model is in the mixed valence
region where the charge fluctuations can not be neglected. To reconcile
the difference in the results one could study the Anderson model in a
regime of parameters that is closer to the Kondo lattice model. This
probably will require the use of a different impurity solver, to test
if the PM-AF transition becomes more weakly first order as the Kondo
limit (larger $U$ and more negative $E_f$) is approached.  The second
question is the relevance of the EDMFT results to real materials and
how the mean field theory should be interpreted.  EDMFT is a mean
field treatment of the spatial degrees of freedom and breaks down in
the immediate vicinity of the transition.  Clearly in the coexistence
region of the upper part of the phase diagram (region I), EDMFT is
unreliable \cite{pankov}. In finite dimensions the system has a
non-trivial anomalous dimension and the transition is second
order. There is a region in the vicinity of the transition where
non-Gaussian thermal fluctuations are important in reality and they
are absent in the EDMFT theory. But outside the vicinity of this
transition we expect the predictions of the theory such as the weak
enhancement of the effective mass to be observable.  Indeed, while
EDMFT induces a spurious first order transition, in the high
temperature regime where the semiclassical evaluation of the free
energy is valid, EDMFT gives a very accurate determination of the
location of the first order phase transition.\cite{pankov,ping1}

The EDMFT solution existed between $J_{c1}(T)$ and $J_{c2}(T)$ lines
at low temperatures (region II) suggests a crossover to a different
regime where non-local effects become important. This has been
recently observed in the YbRh$_2$Si$_2$ system \cite{gegenwart}.  Here
again EDMFT is not reliable in the vicinity of the transition and some
element of non-locality becomes important.  To improve the description
of this region and to reduce the strength of the first order phase
transitions and other deficiencies of EDMFT, in particular the fact
that the spins in the bath are treated classically rather than quantum
mechanically, a different extension of DMFT to treat a cluster of
spins in a self-consistent medium is needed. This line of work will
lead to a two impurity Kondo or Anderson model \cite{jones1,jones2} in
a self-consistently determined bath which is known to have a different
critical point than the spin model in a random magnetic field.

This research was supported by NSF under Grants No. DMR-0096462 and
No. DMR-9983156 and by the Center for Materials Theory at Rutgers
University. We thank P. Coleman, V. Dobrosavljevic, S. Florens,
A. Georges, S. Pankov, and Q. Si for helpful discussions.

\end{document}